\documentclass{IEEEtaes}
\usepackage{amsmath,graphicx, amssymb, comment, derivative, tikz, dsfont,bbm, xcolor}
\usepackage{hyperref}
\usepackage{subcaption}
\usepackage{algorithm}
\usepackage{array}
\usepackage{color,array,amsthm}
\usepackage{textcomp}
\usepackage{stfloats}
\usepackage{url}
\usepackage{verbatim}
\usepackage{graphicx}
\usepackage{cite}
\usepackage[english]{babel}
\usepackage[utf8]{inputenc}
\usepackage[noend]{algpseudocode}
\usetikzlibrary{calc}

\newtheorem{theorem}{Theorem}

\newtheorem{condition}{Condition}

\jvol{XX}
\jnum{XX}
\jmonth{XXXXX}
\paper{1234567}
\pubyear{2022}
\doiinfo{TAES.2022.Doi Number}

\def \states {{X}}
\def \actions {\mathcal{U}}
\def \policy {\boldsymbol{\mu}}
\def \history {\mathcal{I}}
\newcommand{\obj}{{\mathcal{L}}}
\newcommand{\FIM}{{\mathbf{F}}}
\newcommand{\param}{{\theta}}
\newcommand{\paramspace}{{\Theta}}

\newcommand{\ve}{{\mathbf{vec}}}
\newcommand{\tp}{P_{ij}}
\newcommand{\atp}{\mathbf{A}_{iu\mid ju'}}
\newcommand{\cost}{c(i,u)}
\begin{document}

	\title{Fisher Information Approach for Masking the Sensing Plan:  Applications in Multifunction Radars}

	\author{SHASHWAT JAIN}
	\member{Student Member, IEEE}
	\affil{Cornell University, Ithaca, NY, USA} 
	
	\author{VIKRAM KRISHNAMURTHY}
	\member{Fellow, IEEE}
	\affil{Cornell University, Ithaca, NY, USA} 
	
	\author{MURALIDHAR RANGASWAMY}
	\member{Fellow, IEEE}
	\affil{Air Force Research Laboratory}
	
	\author{BOSUNG KANG}
	\member{Member, IEEE}
	\affil{University of Dayton Research
		Institute, Dayton, OH, USA}
	\author{SANDEEP GOGINENI}
	\member{Senior Member, IEEE}
	\affil{Information Systems Laboratories Inc., Dayton,
		Ohio, USA}
	
	\receiveddate{A short version of this paper is accepted for Radar Conference, 2024, under the title: Masking the Sensing Plan for Multifunction Radar in an adversarial setting.}
	
	\corresp{}
	
	\authoraddress{
		S. Jain and V. Krishnamurthy were supported by the US Army Research Office grant W911NF-24-1-0083, Air Force Office of Scientific Research grant FA9550-22-1-0016 and National Science Foundation grant CCF-2312198. M. Rangaswamy and B. Kang were supported by the Air Force Office of Scientific Research (AFOSR) under project 23RYCOR002. S. Gogineni was supported by AFOSR under project 23RYCOR003.}
	
	\editor{}
	\supplementary{}

	\markboth{Jain et al.}{Fisher Information Approach for Masking the Sensing Plan}
	\maketitle

	\begin{abstract}
		
		How to design a Markov Decision Process (MDP) based radar controller that makes small sacrifices in performance to mask its sensing plan from an adversary? The radar controller purposefully minimizes the Fisher information of its emissions so that an adversary cannot identify the controller's model parameters accurately. Unlike classical open loop statistical inference, where the Fisher information serves as a lower bound for the achievable covariance, this paper employs the Fisher information as a design constraint for a closed loop radar controller to mask its sensing plan. We analytically derive a closed-form expression for the determinant of the Fisher Information Matrix (FIM) pertaining to the parameters of the MDP-based controller. Subsequently, we constrain the MDP with respect to the determinant of the FIM. 
		Numerical results show that the introduction of minor perturbations to the MDP's transition kernel and the total operation cost can reduce the Fisher Information of the emissions. Consequently, this reduction amplifies the variability in policy and transition kernel estimation errors, thwarting the adversary's accuracy in estimating the controller's sensing plan.  
	\end{abstract}
	
	\begin{IEEEkeywords}
		Multi-Function Radar, Covert Sensing, Markov Decision Process, Fisher Information Criteria
	\end{IEEEkeywords}
	
	\section{Introduction}
	Markov Decision Processes (MDPs) are widely employed to optimize the sensing plans of sophisticated sensing systems, such as Multi-Function Radars (MFRs) \cite{Haykin2003,haykin2012cognitive, Krishnamurthy2009, Selvi2018}.
	However, a significant vulnerability arises when adversaries observe the state and actions of an MDP. This observation allows them to estimate the underlying model parameters, consequently deducing the sensing plan and potentially taking measures to counteract these decisions.
	
	In this paper, we address the following question: How to design a MDP that makes small sacrifices to its optimization strategy with the aim of masking its sensing plan from an adversary? To achieve this, we propose a novel approach that involves introduction of the Fisher information matrix constraint to the MDP. By doing so, we ensure that the adversary, at best, can only derive a poor-quality estimate. 
	
	In classical statistical inference, Fisher information is employed to characterize the estimation accuracy (covariance) of an open-loop stochastic system. In contrast, this paper utilizes the Fisher information matrix as a design tool for constructing a closed-loop stochastic decision system that effectively conceals its plan from potential adversaries. Another aspect of the paper is that, by incorporating Fisher information matrix constraints, we effectively unify the signal processing layer of a sensing system with the higher-level decision system layer. Traditionally these two layers are designed independently in sensing systems.
	
	
	The paper broadens the scope of Low Probability Intercept Radars (LPIR) beyond their traditional application at the physical layer, where tactics like reducing transmit power aim to confuse passive adversaries such as Radar Warning Receivers (RWRs). Here, we shift towards a decision system-level approach aimed at concealing the radar's sensing plan from passive adversaries, albeit with a trade-off: an increase in total operational cost. These adapted MFRs are termed covert radars, operating discreetly. The sensing plan delineates various operational modes for the MFR. We introduce perturbations to the underlying parameters of the sensing plan, thereby reducing the associated Fisher information. In line with the Cramer-Rao bound (CRB), this reduction in Fisher information corresponds to an increase in the minimum asymptotic error variance for the parameters. The primary objective of this perturbation is to decrease the determinant of the FIM concerning the adversary's estimation of the sensing plan, thereby increasing the minimum asymptotic error variance of the estimate. We regularize the MDP with respect to the determinant of the FIM. We demonstrate that perturbing the conditional transition matrix and state-action costs of the underlying MDP governing the mode transition of the radar controller leads to a higher error in estimating the sensing plan. Drawing from the principles Information Geometry, we utilize the determinant of the Fisher Information due to its ability to gauge the curvature of the likelihood function. Lower curvature of the likelihood function correlates with a lower volume, resulting in higher variance of the parameter estimate for the adversary.
	\subsection*{Related Works}
	MFRs were introduced in \cite{Haykin2003} to address the requirement of scenario specific adaptivity for the modern radars. An extensive survey of scenario based MFR operations has been documented in \cite{klemm2017novel1,klemm2017novel2}, where the switching of the operation parameters like waveform, pulse length, carrier frequency, polarization, pulse repetition frequency (PRF) and dwell time, is demonstrated to adapt to various scenarios. In \cite{Visnevski2007}, the MFR Mercury's modes of operation are modeled using Stochastic Context-Free Grammars (SCFG) and Markov Modulated Markov Chains. In \cite{Wang2008}, task scheduling for the MFR Mercury is analyzed using SCFG. In \cite{bao2023online}, the modes of an MFR are modeled using Hidden Markov Models. In \cite{Krishnamurthy2009,Charlish2020}, Partially Observed Markov Processes (POMDP) are employed for target search and track management. In \cite{Wintenby2006}, a radar task scheduler is proposed, utilizing a hierarchical cross-layer MDP. The paper highlights the unreliability of scheduling performance of the heuristic scheduling-based radar controllers as they do not optimize operation costs. In contrast, MDP-based controllers leverage dynamic programming to effectively optimize the operation  cost. In the work presented in \cite{Selvi2018}, reinforcement learning (RL) is employed for radar systems to compute the optimal policy in diverse scenarios. This involves tracking multiple targets amidst varying interference patterns. The paper establishes a detailed modeling of Signal-to-Interference-plus-Noise Ratio (SINR)-based state-action costs for the MFR. 
	
	A substantial body of literature addresses the masking of radar sensing plans at the physical layer level like \cite{Blair1998}. Recent investigations in the decision systems level include studies on intelligent sensor selection for beam allocation in a phased array radar under passive adversarial conditions \cite{Dai2020}. In \cite{JIANG2023103952}, anti-jamming decision methods for a radar network employ multi-agent deep reinforcement learning, developing a deterministic policy to mitigate jamming signals. \cite{Thornton2021} proposes a constraint online learning algorithm for optimal waveform selection to counter intentional adaptive jamming. \cite{Howard2022} models radar-adversarial interactions as a multi-armed adversarial bandit problem, demonstrating the efficiency of adversarial multi-player multi-armed bandit algorithms (MMAB) in maintaining target tracking precision despite introduced emissions. \cite{kang2023} treats the radar-adversary interaction as a game, proposing a method to design an interference signal that confuses the adversary radar while minimizing interference power. This ensures the signal-to-clutter-plus-noise ratio (SCNR) of the radar stays below a specific threshold. For a comprehensive overview of adversary mitigation strategies in radars, \cite{Zhang2023} provides an extensive review. 
	
\begin{figure}
		\centering
		\scalebox{0.8}{
			\begin{tikzpicture}
				\node[draw, fill=blue!20, minimum width=7.5cm, minimum height=3.5cm] (MFR) at (0,0) {};
				\node (LabelMFR) at (0,1) {MFR Controller};
				\node[draw, fill=green!20, minimum width=3cm, minimum height=2cm, align=center, text width=2cm] (MDP) at (-2,-0.5) {Fisher Information\\Constrained MDP};
				\node[draw, fill=red!20, minimum width=2cm, minimum height=2cm, align=center, text width=2cm] (Sensing) at (2,-0.5) {Masked\\Sensing Plan\\($\mu$)};
				\node[draw, dashed, minimum width=7.5cm, minimum height=2.5cm, violet] (Environment) at (0,-4.5) {};
				\node (LabelEnv) at (0,-6) {Environment};
				\node[draw, fill=orange!20, minimum width=7cm, minimum height=2cm, align=center] (Adversary) at (0,-4.5) {Adversary with Likelihood function $\mathcal{L}$};
				\draw[->,thick] (MDP.east) -- (Sensing.west);
				
				\node[above] (yk) at (3.1,-2.95) {$y_{k}\sim\mu$};
				\foreach \y in {-1.75,-2.00,...,-3.25} {
					\draw (1.75,\y) arc[start angle=-135, end angle=-45, radius=0.5cm];
				}
				
				\draw[->,thick,dotted, red] ( Adversary.north -| MDP.south) -- (MDP.south);
				\node (MLE) at (-2.5,-2.75) {$\mathcal{L}$};
			\end{tikzpicture}
		}
		\caption{The Fisher Information-based masked sensing approach relies on leveraging the Likelihood function (red arrow), $\mathcal{L}$, of the adversary to solve a constrained MDP optimization problem. The sensing plan $\mu$ is computed by minimizing the Fisher information for the adversary at the cost of perturbing the parameters of the MDP. Consequently, the sensing samples $y_k$ obtained from the masked sensing plan (illustrated by the arcs extending from the MFR to the Adversary) provide an inaccurate estimate of the MFR controller's MDP underlying parameters to the adversary.}
		\label{fig:MainIdea}
	\end{figure}
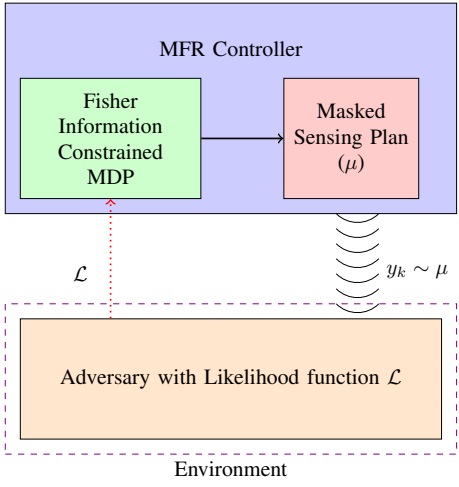

	The challenge of masking of the sensing plan in an adversarial setting has arisen across multiple domains. In wireless communications, a trade-off between spoofing and jamming has been investigated for cognitive radio, and venerability in sensing of a cognitive radio network are exploited in \cite{Peng2011}. In \cite{Zhang2020}, the attacker masks its attack strategy by employing the Kullback-Leibler (KL) divergence criteria to minimize the attacking cost while maximizing the detection cost for the controller. In \cite{Karabag2022}, the controller perturbs its actions to deceive the adversary to satisfy a state reachability condition. In \cite{Pattanayak2022}, the radar masks its objective function using Inverse-Inverse Reinforcement Learning (IIRL) in an adversarial presence. The approach is non-parametric and relies on the fact that the adversary is using Afriat's theorem to estimate the radar's cost of operation. The radar perturbs its responses in order to increase the max-margin of the adversary's cost estimate. The approach does not highlight the variance of the cost estimated by the adversary, unlike this paper, where we leverage the Fisher information criteria in order to increase the lower bound for the variance of the adversary's parameter estimate.  In \cite{jain2023controlling}, the utilization of reinforcement learning in an adversarial setting is demonstrated for masking the optimal plan while simultaneously minimizing an objective function. In \cite{Savas2020}, a random policy is synthesized using the maximum entropy criteria to increase the entropy of the resulting MDP. In \cite{Karabag2019}, a masking strategy is proposed, leveraging the minimization of Fisher information associated with the average time spent in each state to ensure the satisfaction of a state reachability constraint. The conditional transition matrix and the costs are known both to the controller and the adversary, the controller perturbs its actions to mislead the adversary. Diverging from \cite{Karabag2019}, our approach involves reducing the determinant of the Fisher information for the augmented state-action Markov chain through the randomization of the sensing plan/optimal policy. In our study, the adversary lacks knowledge of the state transition kernel conditioned on actions and costs. To the best of our knowledge, our masking approach is novel. We illustrate the Fisher information based masked sensing for MFR in Fig.\ref{fig:MainIdea}.
	\subsection*{Organization and Main Results}
	\begin{enumerate}
		\item In Sec.\,\ref{Sec:GO}, we review the MDP based sensing plan for a radar controller in the absence of an adversary and derive the expression for the determinant of the FIM for a Markov Decision Process. Additionally, we establish the regularity conditions necessary for the existence of the inverse of Fisher information matrix of the finite state finite action MDP employed by the radar controller for mode switching operations.
		\item In Sec.\,\ref{Sec:ECCM}, we constrain the MDP by the determinant of the FIM, to mask the sensing plan. Our formulation incorporates the logarithm of the determinant of the Fisher Information Matrix (FIM) as a constraint for the MDP. This introduces a trade-off involving perturbations in the total cost of mode switching operations, the conditional transition matrix, and the state-action costs. 
		\item In Sec.\,\ref{Sec:RadJamInt}, we illustrate a comprehensive model outlining state-action costs and the conditional transition matrix. We integrate the parameters of the physical layer of the MFR with the higher-level decision layer.
		\item Section\,\ref{Sec:Num} furnishes numerical examples involving a 10-state, 4-action radar controller. We illustrate the reduction in the determinant of the Fisher information matrix in cases when we perturb: a) the total cost of operation, b) the state-action costs and c) the conditional transition matrix, due to the introduction of the nonlinear Fisher information constraint. We empirically demonstrate that the Fisher Information yields a higher error in the estimate of the conditional transition matrix  compared to the Maximum Entropy criteria\,\cite{Savas2020}, and is therefore a better criteria to mask the sensing plan.
	\end{enumerate}

	\section{Fisher Information for a Markov Decision Process}
	\label{Sec:GO}
	
	In this section we introduce the sensing plan utilized by a MDP based radar controller and the associated Fisher Information for the adversary. The sensing plan is the optimal policy for MDP used to minimize the total cost of operation. The MDP based sensing plan for the radar controller is introduced in Sec.\,\ref{Sec:GO}-\ref{Sec:MDP}. In Sec.\,\ref{Sec:GO}-\ref{Sec:FIM}, we derive an expression for the determinant of the FIM for a MDP and introduce the conditions for the existence of the inverse of FIM for MDP.

	\subsection{MDP based Sensing Plan}
	\label{Sec:MDP}
	In this section we introduce the MDP based mode switching for the controller. The switching is modelled using an MDP, as defined in \cite{krishnamurthy2016}. A MDP consists of
	\begin{enumerate}
		\label{sec:defMDP}
		\item A finite state space $\states$.
		\item A finite action space $\actions$.
		\item Conditional transition matrix $\tp(u)$ for $i, j \in \states$ and $u \in \actions$, the radar controller's mode switching probability.
		\item State-action transition costs $\cost,\,i\in\states,\,u\in\actions$, instantaneous costs incurred while operating the radar.
	\end{enumerate}
	In Sec.\,\ref{Sec:RadJamInt}, we will demonstrate how $\tp(u)$ and $\cost$ are connected with the physical layer parameters of the radar.
	
	The \textit{history} $\history_{k}$ at time $k$ is a sequence of states and actions such that $\history_{k} = x_{0} u_{0} x_{1} \ldots x_{k-1} u_{k-1} x_{k}$ where $\{x_k\}\in\states$ and $\{u_k\}\in\actions$. A \textit{policy} is a sequence $\policy = \mu_0\mu_1\ldots$ where each $u_k= \mu_k(\history_k)$ is a mapping from $\history_k$ to $\actions$.   A \textit{stationary policy} is a sequence $\policy = \mu \mu \ldots$ where $u_k=\mu_{k}(x_k)$. The optimal policy is the radar controller's sensing plan.
	
	The radar controller optimizes the infinite horizon average cost expressed as
	\begin{equation}
		\label{eq:cost}
		\underline{V}_{\policy}(i)=\lim_{N\rightarrow\infty}\frac{1}{N+1}\mathbb{E}_{\policy}\bigg\{\sum_{k=0}^{N}c(x_k,u_k)|x_0=i\bigg\},
		i\in\states.
	\end{equation}
	The optimal policy is \begin{equation}
		\label{eq:optpolicy}
		\policy_{0}=\underset{\policy}{\text{argmin}}\,\underline{V}_{\policy}(i).
	\end{equation}
	Since the controller is solving the infinite horizon problem, the initial decisions do not influence the optimal cost. As a result, it is intuitive that the policy $\mu^{*}$ is also optimal at subsequent times $k=1,2,\dots$. and we obtain a stationary policy, $\policy=\{\mu,\mu,\dots\}$. 
	
	Equation \eqref{eq:optpolicy} is solved using the following linear program:
	\begin{align}
		\label{eq:LPPrim}
		\boldsymbol{\pi}_{0}=\underset{\pi}{\text{argmin}}&\quad\sum_{i\in\states}\sum_{u\in\actions}c_0(i,u)\pi(i,u)\\
		\text{s.t.}&\quad\pi(i,u)\geq 0,\quad i\in\states,\,u\in\actions\nonumber\\
		&\sum_{u}\pi(j,u)=\sum_{i}\sum_{u}\tp^{0}(u)\pi(i,u),\,j\in\states\nonumber\\
		&\sum_{i}\sum_{u} \pi(i,u)=1\nonumber
	\end{align}
	where $c_0(i,u)$ is the state-action transition cost, $\tp^{0}(u)$ is the conditional transition matrix when no adversary is present. 
	The objective in problem \eqref{eq:LPPrim} is the total cost of operating the radar contrtoller. The optimal policy or the sensing plan is 
	\[\mu_0(i)=u\quad\text{w.p}\quad\frac{\pi_0(i,u)}{\sum_{u}\pi_0(i,u)},\quad i\in\states\]
	Note that the sensing plan is deterministic over the actions  for a given state even though it appears to be randomized.
	
	The adversary infers the sensing plan $\mu$ by observing the states and actions of the induced Markov process. In the next section we define the Fisher information matrix for the estimate of the transition matrix induced by the sensing plan.

	\subsection{Fisher Information for an MDP}
	\label{Sec:FIM}
	
	In this section, we derive an expression for the determinant of the FIM for a MDP, \eqref{eq:detFIM1}. We delineate the conditions under which the inverse of Fisher information of the adversary's estimate of the transition probability of the MDP-based radar controller exists. The expression for the FIM will be incorporated as a constraint for the sensing plan in the subsequent section, with the goal of reducing the adversary's Fisher information. For a given action space $u\in\actions$ and state space $i\in\states$ the state-action augmented transition matrix represents a $|\states||\actions|$ state Markov Chain. The stationary augmented state action transition matrix, denoted as $\atp$, observed by the adversary is:
	\begin{align*}
		&\atp=P(x_{k+1}=j,\,u_{k+1}=u{'}\mid x_{k}=i,\,u_{k}=u)\\ &=P(u_{k+1}=u^{'}\mid x_{k+1}=j) P(x_{k+1}=j\mid x_{k}=i,\,u_{k}=u)
	\end{align*}
	Denoting $P(u_{k+1}=u{'}, x_{k+1}=j)=\pi(j,u)$ and $P(x_{k+1}=j\mid x_{k}=i,\,u_{k}=u)=\tp(u)$, we have:
	\begin{align}
		\label{eq:AugTrans}
		\atp=\frac{\pi(j,u{'})}{\sum_{u}\pi(j,u)}\tp(u) 
	\end{align}
	
	Let $\{y_k\}$ be the tuple denoting the joint state-action $\{x_k,\,u_k\}$, $m$ and $n$ denote the indices for the state action pair.  We define the maximum likelihood estimator for the adversary, as illustrated in Fig.\,\ref{fig:MainIdea}, for the stationary transition matrix $\mathbf{A}^{\param}$ parameterized by $\param=\{\theta_1,\dots,\theta_r\}\in\paramspace$ as:
	\begin{align}
		\label{eq:MLE}
		\mathcal{L}_{N}({\param})=\sum_{k=1}^{N}\log p(y_{k+1},\,y_{k};\,\param),
	\end{align}
	where 
	\begin{equation*}
		p(y_{k+1},\,y_{k};\,\param)=\prod_{m,n}^{|\states||\actions|} (a_{mn}^{\param})^{\mathbbm{1}(\{y_{k+1}=m,\,y_{k}=n\})},
	\end{equation*}
	$\mathbbm{1}(\cdot)$ is the indicator function and $a_{mn}^{\param}=[\mathbf{A}^{\param}]_{mn}$. Note that $r\leq|\states||\actions|(|\states||\actions|-1)$, since $\mathbf{A}$ is a Markov chain such that $\sum_{n}a_{mn}=1$. Denoting the parameter for the adversary to be $\param_{ad}$, the maximum likelihood estimate of $\mathbf{A}_{\param_{ad}}$ is the solution of the equation:
	\begin{equation}
		\label{eq:MLESol}
		\partial_{\param_{\alpha}}\mathcal{L}_{n}(\param)\Bigr|_{\param=\param_{ad}}=0,\quad1\leq\alpha\leq r.
	\end{equation}
	The FIM, $[\FIM(\param_{ad})]_{\alpha\beta}=\mathbb{E}_{\param_{ad}}\big[\partial_{\param_{\alpha}}\mathcal{L}_{n}(\param)\partial_{\param_{\beta}}\mathcal{L}_{n}(\param)\big]$, which for the transition kernel of the Markov chain is:
	\begin{equation*}
		\label{eq:FIM}
		[\FIM(\param_{ad})]_{\alpha\beta}=-\sum_{mn}^{\states\times\actions}a_{m}^{\param_{ad}}a_{mn}^{\param_{ad}}\big(\partial_{\param_{\alpha}}{\log\,a_{mn}^{\param}}\big)\big(\partial_{\param_{\beta}}{\log\,a_{mn}^{\param}\big)}
	\end{equation*}
	where $1\leq \alpha,\beta\leq r$, $a_{m}^{\param_{ad}}$ is the stationary distribution for the transition matrix $\mathbf{A}_{\param_{ad}}$. A simplified notation for the Fisher Information matrix is given in \cite{Takeuchi2009}, given that $1\leq n\leq |\states||\actions|-1$:
	
	\begin{align*}
		\FIM_{(mn),(m'n')}=\delta(m,m')a_{m}^{\param_{ad}}\Bigg(\frac{\delta(n,n')}{a_{mn}^{\param_{ad}}}+\frac{1}{a_{1n}^{\param_{ad}}}\Bigg),
	\end{align*}
	where $\delta(\cdot,\cdot)$ is the Kronecker delta function and $a_{m}^{\param_{ad}}=\frac{\Delta_{mm}}{\sum_{m}\Delta_{mm}}$ is the stationary distribution of the augmented state action transition matrix $\mathbf{A}_{\param_{ad}}$, $\Delta_{mm}$ being the cofactor of the $(m,m)^{\text{th}}$ element of $\mathbf{I}-\mathbf{A}_{\param_{ad}}$ matrix, where $\mathbf{I}$ is the identity matrix. The determinant of the Fisher Information matrix is:
	\begin{equation}
		\label{eq:detFIM}
		\det(\FIM(\param_{ad}))=\prod_{m\in\states\times\actions}\frac{(a_{m}^{\param_{ad}})^{|\states||\actions|}}
		{\prod_{n\in\states\times\actions}a_{mn}^{\param_{ad}}}.
	\end{equation}
	In order for the determinant of the Fisher information matrix in \eqref{eq:detFIM} to be defined, each element of the augmented state-action transition matrix $\mathbf{A}$ should be positive. This ensures the determinant of the Fisher Information to be finite. This can only happen if each element of the conditional transition matrix $\tp(u)$ is positive{\footnote{The condition that all elements of $\tp(u)>0$ is quite strong, however if elements in $\tp(u)$ are zero, we decompose the resulting state-action Markov chain into recurrent classes where $\tp(u)>0$ and compute FIM separately for each recurrent class. For the sake of simplicity we assume $\tp(u)>0$.}}. If all elements of $\tp(u)$ are positive then the resulting Markov chain $\mathbf{A}$ is unichain\footnote{\cite{krishnamurthy2016}, An MDP is said to be unichain if every deterministic stationary policy yields a Markov chain with a single recurrent class and a set of transient states which maybe empty.}. We observe that $\param_{ad}$ depends only $\pi(i,u)$ and $\tp(u)$, which we will optimize in the subsequent sections with respect to costs $\cost$ and $\tp(u)$ in order to minimize the Fisher information for the adversary.
	Observe that for the stationary policy we have the following relation:
	
	\begin{align*}
		&P(x_{k+1}=j,\,u_{k+1}=u',\,x_{k}=i,\,u_{k}=u)\\
		&=\mathbf{A}_{iu\mid ju'}P(x_{k}=i,\,u_{k}=u)\\
		&=\mathbf{A}_{ju'\mid iu}P(x_{k+1}=j,\,u_{k+1}=u')
	\end{align*}
	Since the policy is stationary, we have $P(x_{k+1}=j,\,u_{k+1}=u')=\pi(j,u')$ and $P(x_{k}=i,\,u_{k}=u)=\pi(i,u)$. So, we have:
	\begin{equation*}
		\atp\pi(i,u)=\mathbf{A}_{ju'\mid iu}\pi(j,u').
	\end{equation*} 
	Thus augmented state-action matrix $\mathbf{A}_{mn}$ is a reversible Markov chain with $\pi(i,u)$ is the stationary distribution, as it establishes detailed balance. The unichain condition implies that the resulting Markov chain induced by the stationary policy $\mu$ is recurrent for all states leading to $a_{m}^{\param}=\pi(i,u)$, where $m$ represents the index for the tuple $\{i,u\}$. In that case the the determinant of the FIM is:
	\begin{equation*}
		\label{eq:detFIMsimple}
		\det(\FIM)=\prod_{i\in\states,u\in\actions}\frac{\pi(i,u)^{|\states||\actions|}}{\prod_{j\in\states,u'\in\actions}\frac{\pi(j,u{'})}{\sum_{u}\pi(j,u)}\tp(u)}
	\end{equation*}
	Taking the logarithm we have:
	\begin{align*}
		&\log\det\FIM=|\states||\actions|\sum_{i\in\states,u\in\actions}\log\pi(i,u)\\
		&-\sum_{i\in\states,u\in\actions}\sum_{j\in\states,u'\in\actions}\log\pi(j,u')\nonumber\\
		&+\sum_{i\in\states,u\in\actions}\sum_{j\in\states,u'\in\actions}\log\sum_{u}\pi(j,u)\nonumber\\
		&-\sum_{i\in\states,u\in\actions}\sum_{j\in\states,u'\in\actions}\log \tp(u)\nonumber
	\end{align*}
	which can be further simplified as:
	\begin{align*}
		&\log\det\FIM=|\states||\actions|^2\sum_{j\in\states}\log\sum_{u\in \actions}\pi(j,u)\\
		&-|\actions|\sum_{i,j\in\states,u\in\actions}\log \tp(u)\nonumber
	\end{align*}
	The expression for the determinant of the FIM is:
	\begin{equation}
		\det\FIM=\frac{\prod_{i\in\states}\Bigg(\sum_{u\in\actions}\pi(i,u)\Bigg)^{|\actions|^2|\states|}}{\prod_{i,j\in\states,u\in\actions}\tp(u)^{|\actions|}}
		\label{eq:detFIM1}
	\end{equation}
	It's important to note that $\pi$ is implicitly dependent on both the state-action transition costs $\cost$ and the conditional transition matrix $\tp(u)$. For brevity, we leave the expression for the FIM in terms of $\pi$ and $\tp(u)$. The parameters that determine the FIM for the adversary are expressed using only two variable $\pi$ and $\tp(u)$, i.e., $\param_{ad}=\{\pi,\tp(u)\}$, with the adversary aiming to estimate $\param_{ad}$.
	
	The likelihood function \eqref{eq:MLE} satisfies the regularity conditions \cite[Condition 1.1, 1.2]{Billingsley1961}, which ensure that the inverse of the FIM exists. We invoke \cite[Thm. 5.1]{Billingsley1961} to show that \eqref{eq:MLE} is a consistent estimator and the estimate $\hat{\param}_{ad}$ converges in law to a normal distribution characterized by the FIM. 
	\begin{condition}[\cite{Billingsley1961}, Cond.\,5.1]
		There exists a set $D$ of $(m,n)\in\states\times\actions$ such that $a_{mn}(\param_{ad})>0$ is independent of $\param_{ad}$ and each $a_{mn}(\param_{ad})$ has continuous partial derivatives of third order throughout $\paramspace$. Moreover the $d\times r$ matrix 
		$\partial_{\param_{\alpha}}a_{mn}(\param)\quad (m,n)\in D, \alpha=1,\cdots,r,$
		($d$ being the number of elements in the set $D$) has rank $r$ throughout $\paramspace$. For each $\param_{ad}$ there exists only one ergodic set fo states and there are no transient states.
	\end{condition} 
	\begin{theorem}[\cite{Billingsley1961}, Thm.\,5.1]
		Suppose that $[y_{k},A_{\param}, \paramspace]$, satisfies Cond.\,1. Then there exists a sequence $\{\hat\param_{ad}\}$ of random vectors in $\paramspace$, each one being a function $\hat{\param}_{ad}=\hat{\param}_{ad}(y_1,\cdots,y_{N+1})$ of the observation, such that $\hat{\param}_{ad}$ converges in probability to the true $\param_{ad}$, defined in \eqref{eq:MLESol}. The $\hat{\param}_{ad}$ is a solution of \eqref{eq:MLESol} with probability going to one as $N\rightarrow\infty$. Thus \eqref{eq:MLE} is a statistical consistent maximum-likelihood estimator of $\param_{ad}$. Moreover $\hat{\param}_{ad}$ is a local maximum of $L_{N}(\param)$ with probability going to one. Finally, if $\bar{\param}_{ad}$ is a second consistent solution of \eqref{eq:MLESol}, then the probability that $\hat{\param}_{ad}=\bar{\param}_{ad}$ goes to one as $N\rightarrow\infty$. Further,
		\begin{align}
			\label{eq:CLT}
			\sqrt{N^{-1}}({\hat\param}_{ad}-\param_{ad})&\xrightarrow{\text{in law}}\mathcal{N}(0,\FIM^{-1}(\param_{ad})),
		\end{align}
		where $\mathcal{N}(\cdot,\cdot)$ is a Normal distribution.
	\end{theorem}
	
	By \eqref{eq:CLT}, \eqref{eq:MLE} is an asymptotically unbiased estimator, so by \cite{kay1993fundamentals}, the covariance matrix of the estimate and Fisher information matrix are related using the following inequality: 
	\begin{equation}
		\label{eq:CRB}
		\mathbf{Cov}({\hat{\param}}_{ad})-\FIM^{-1}(\param_{ad})\geq 0
	\end{equation}
	where $\mathbf{A}-\mathbf{B}\geq 0$ means $\mathbf{A}-\mathbf{B}$ is a positive semi-definite matrix and $\mathbf{Cov}$ is the covariance matrix for the estimate ${\hat{\param}}_{ad}$. Eq.\,\eqref{eq:CRB} states that the lowest bound of the variance of the estimate of the adversary's transition matrix is given by the inverse of the Fisher information, this provides us the rationale to reduce the determinant of the FIM of the adversary. In \eqref{eq:detFIM1}, we have defined $\det\FIM(\param_{ad})$ as a function of the sensing plan $\mu$.
	
	\section{Minimizing  Fisher Information to Mask Sensing Plan}
	\label{Sec:ECCM}

	In this section, we introduce the Fisher information criteria as a constraint in the MDP based masked sensing plan. We demonstrate that the introduction of the Fisher information constraint the sensing plan is masked at the expense of perturbation in (1) the total operation cost (Sec.\,\ref{Sec:ECCM}-\ref{Sec:TCostPert}); (2) the state action cost (Sec.\,\ref{Sec:ECCM}-\ref{Sec:StateActionCost}) and (3) the conditional transition matrix (Sec.\,\ref{Sec:ECCM}-\ref{Sec:CondTMat}).
	
	As illustrated in Fig.\ref{fig:MainIdea}, an adversary observes the radar controller and infers actions and states of the radar controller to obtain $\history_k$. The data $\history_k$ is sequence of $y_k$ which is sampled from $\mu$, where $\mu$ represents the sensing plan for the MDP, \eqref{eq:LPPrim}. The adversary uses an unbiased MLE ($\mathcal{L}$), \eqref{eq:MLE}, to estimate the sensing plan by observing the state-action trajectory. The controller aims to devise a masking plan such that the variance in the error of the estimate is large. The controller needs to increase the determinant of the parameter error covariance matrix, thereby decreasing the determinant of the FIM of the adversary. The controller perturbs the optimal cost computed using \eqref{eq:LPPrim} by introducing an additional Fisher information constraints. Now we formulate different instances of the FIM determinant minimization problems for the adversary.
	
	\subsection{Masking the Sensing Plan by Perturbing the Total Operation Cost}
	\label{Sec:TCostPert}
	Recall that the objective of \eqref{eq:LPPrim} is the total cost of operating the radar where the resultant sensing plan or the optimal policy is deterministic. We perturb the total cost of operation in order to minimize the Fisher information for the adversary. This is achieved by introducing the FIM constraint as a global constraint which randomizes the sensing plan thereby masking it. 
	The optimization problem is:
	\begin{align}
		\label{eq:ECCMLog}
		\underset{\pi}{\min}&\quad\Bigg(\sum_{i}\sum_{u}c_0(i,u)(\pi(i,u)-\pi_{0}(i,u))\Bigg)^2\\
		\text{s.t.}&\quad\pi(i,u)\geq 0,\quad i\in\states,\,u\in\actions\nonumber\\
		&\sum_{u}\pi(j,u)=\sum_{i}\sum_{u}\tp^{0}(u)\pi(i,u),\,j\in\states\nonumber\\
		&\sum_{i}\sum_{u} \pi(i,u)=1\nonumber\\
		&\det\FIM(\pi)\leq \det\FIM(\pi_0)\nonumber
	\end{align}
	
	$\tp^{0}(u)$, $c_0(i,u)$ and $\pi_0$ are defined in \eqref{eq:LPPrim}. $\det\FIM(\cdot)$ is defined in \eqref{eq:detFIM1}. The objective function in \eqref{eq:ECCMLog} is perturbation in the cost of operation. Since we are not perturbing the transition matrix $\tp(u)$, the Fisher information is dependent only on the policy $\pi$. As stated in Sec.\,\ref{Sec:GO}-\ref{Sec:MDP}, the optimal policy for \eqref{eq:LPPrim} is deterministic, implying $\pi_{0}(i,u)=0$ for some $i\in\states,\,u\in\actions$, so $\det\FIM(\pi_0)$ is infinite. We reformulate \eqref{eq:ECCMLog} below, with $\gamma$ being the Lagrange multiplier for the Fisher information constraint:
	\begin{align}
		\label{eq:ECCMLog1}
		\underset{\pi}{\min}&\quad\Bigg(\sum_{i}\sum_{u}c_0(i,u)(\pi(i,u)-\pi_{0}(i,u))\Bigg)^2\nonumber\\
		&+\gamma\log \Bigg(\prod_{i\in\states}\sum_{u\in\actions}\pi(i,u)\Bigg)\\
		\text{s.t.}&\quad\pi(i,u)\geq 0,\quad i\in\states,\,u\in\actions\nonumber\\
		&\sum_{u}\pi(j,u)=\sum_{i}\sum_{u}\tp^{0}(u)\pi(i,u),\,j\in\states\nonumber\\
		&\sum_{i}\sum_{u} \pi(i,u)=1\nonumber.
	\end{align}
	The masked sensing plan is $\mu_{*}(i)=u\quad\text{w.p}\quad\frac{\pi(i,u)}{\sum_{u}\pi(i,u)}.$
	Equation \eqref{eq:ECCMLog1}, represents the simultaneous minimization of total operation cost perturbation and the determinant of FIM of the resulting augmented state-action transition matrix defined in \eqref{eq:AugTrans}. The objective in \eqref{eq:ECCMLog1} is the perturbation in the total operating cost along with determinant of the Fisher information introduced as a regularization. 
	Note that \eqref{eq:ECCMLog1} is a non-convex optimization problem.

	\subsection{Masking the Sensing Plan by Perturbing the State-Action Cost}
	\label{Sec:StateActionCost}
	We introduce additional perturbations to the cost in \eqref{eq:ECCMLog1} to reduce the Fisher Information for the adversary. In this case we include an additional perturbation in the state-actions cost term in the objective function. So, \eqref{eq:ECCMLog1} becomes 
	\begin{align}
		\label{eq:ECCMLogCost}
		\underset{\pi, \cost}{\min}&\quad\Bigg(\sum_{i}\sum_{u}\cost\pi(i,u)-c_{0}(i,u)\pi_{0}(i,u)\Bigg)^2\nonumber\\&+\gamma_1\sum_{i}\sum_{u}(\cost-c_0(i,u))^2\nonumber\\&+\gamma_2\log \Bigg(\prod_{i\in\states}\sum_{u\in\actions}\pi(i,u)\Bigg)\\
		\text{s.t.}&\quad\pi(i,u)\geq 0,\quad i\in\states,\,u\in\actions\nonumber\\
		&\sum_{u}\pi(j,u)=\sum_{i}\sum_{u}\tp^{0}(u)\pi(i,u),\,j\in\states\nonumber\\
		&\sum_{i}\sum_{u} \pi(i,u)=1\nonumber
	\end{align}
	where the masked sensing plan is: $\mu_{*}(i)=u\quad\text{w.p}\quad\frac{\pi(i,u)}{\sum_{u}\pi(i,u)}$.
	$\tp^{0}(u)$, $c_0(i,u)$ and $\pi_0(i,u)$ are defined in \eqref{eq:LPPrim}, $\cost$ is the perturbed state state-action cost. $\gamma_1$ and $\gamma_2$ are Lagrange multipliers which penalize the state-action cost perturbation and the Fisher information, respectively. The term $\sum_{i\in\states,u\in\actions}(\cost-c_{0}(i,u))^2$ in \eqref{eq:ECCMLogCost} is a bound on the norm of the stat-action cost penalized by $\gamma_1$. The objective function in \eqref{eq:ECCMLogCost} is perturbation in the cost of operation, along with perturbation of state-action costs and the Fisher information constraint introduced as a regularization.
	In \eqref{eq:ECCMLogCost}, both the policy $\pi$ and the state-action costs $\cost$ are computed. To solve \eqref{eq:ECCMLogCost}, we employ block co-ordinate descent algorithm demonstrated in Algorithm \ref{alg:BCD}. 
	\begin{algorithm}
		\caption{Block Co-ordinate Descent Algorithm to compute State-Action Costs and Optimal Policy in \eqref{eq:ECCMLogCost}}\label{alg:BCD}
		\begin{algorithmic}[1]
			\State \textbf{Input:} $\gamma_1$, $\gamma_2$, $\tp(u)$, $c_{0}(i,u)$ and $\pi_{0}(i,u)$
			\State \textbf{Initialize:} $c_{in}(i,u)$, $\pi_{in}(i,u)$
			\State \textbf{Define:} $\boldsymbol{\varphi}_{\pi}(c):=\Bigg(\sum_{i}\sum_{u}\cost\pi(i,u)-c_{0}(i,u)\pi_{0}(i,u)\Bigg)^2+\gamma_1\sum_{i}\sum_{u}(\cost-c_0(i,u))^2$
			\State \textbf{Define:} $\boldsymbol{\varphi}_{c}(\pi):=\Bigg(\sum_{i}\sum_{u}\cost\pi(i,u)-c_{0}(i,u)\pi_{0}(i,u)\Bigg)^2+\gamma_2\log \Bigg(\prod_{i\in\states}\sum_{u\in\actions}\pi(i,u)\Bigg)$
			\State \textbf{Define:} $\mathcal{S}=\{\pi: \quad\pi(i,u)> 0\,\text{and}\,\sum_{u}\pi(j,u)=\sum_{i}\sum_{u}\tp(u)\pi(i,u),\,j\in\states\,\text{and}\,\sum_{i}\sum_{u} \pi(i,u)=1\}$
			\While{$\epsilon_{k-1}>\text{Threshold}$}
			\State $c_k(i,u)\gets\underset{c}{\text{argmin}}\,\boldsymbol{\varphi}_{\pi_{k-1}}(c)$
			\State $\pi_k(i,u)\gets\underset{\pi\in\mathcal{S}}{\text{argmin}}\,\boldsymbol{\varphi}_{c_{k}}(\pi)$
			\State $\epsilon_k\gets \Vert([c_k;\pi_k]-[c_{k-1};\pi_{k-1}])\Vert_{2}^{2}$
			\EndWhile\label{euclidendwhile}
			\State \textbf{return} $[c;\,\pi]$\Comment{The state-action cost and the optimal policy}
		\end{algorithmic}
	\end{algorithm}
	
	\subsection{Masking the Sensing Plan by Perturbing the Conditional Transition Matrix}
	\label{Sec:CondTMat}
	In this case, the radar perturbs the conditional transition kernel $\tp^{0}$ in order to minimize the Fisher information of the adversary. We solve the following optimization problem:
	\begin{align}
		\label{eq:ECCMLogTransition}
		\underset{\pi, \tp(u)}{\min}&\quad\Bigg(\sum_{i}\sum_{u}c_0(i,u)(\pi(i,u)-\pi_{0}(i,u))\Bigg)^2\nonumber\\&+\gamma_1\sum_{u}\lVert \ve(P^{T}(u))-\ve(P_{0}^{T}(u))\rVert_{1}\nonumber\\&+\gamma_2\log\Bigg(\frac{\prod_{i\in\states}(\sum_{u}\pi(i,u))^{|\states||\actions|}}{\prod_{i,j\in\states,u\in\actions}\tp(u)}\Bigg)\\
		\text{s.t.}&\quad\pi(i,u)\geq 0,\quad i\in\states,\,u\in\actions\nonumber\\
		&\sum_{u}\pi(j,u)=\sum_{i}\sum_{u}\tp(u)\pi(i,u),\,j\in\states\nonumber\\
		&\sum_{i}\sum_{u} \pi(i,u)=1\nonumber\\
		&\tp(u)> 0\quad i,j\in\states,\,u\in\actions\nonumber\\
		&\sum_{j}\tp(u)=1\quad i\in\states,\,u\in\actions\nonumber
	\end{align}
	where the masked sensing plan is: $\mu_{*}(i)=u\quad\text{w.p}\quad\frac{\pi(i,u)}{\sum_{u}\pi(i,u)}$. $\tp^{0}(u)$, $c_{0}(i,u)$ and $\pi_{0}(i,u)$ are defined in \eqref{eq:LPPrim}. For transition matrix perturbation, we are bounding the sum of total variation distance between each row of the conditional transition matrix for all actions, which is characterized by the Lagrange multiplier corresponding to $\gamma_1$ in \eqref{eq:ECCMLogTransition}. The Lagrange multiplier corresponding to the FIM is denoted using $\gamma_2$. $\mathbf{vec}(\cdot)$ is the vectoring operation w.r.t to each row of the conditional transition matrix $\tp(u)$.
	The objective function in \eqref{eq:ECCMLogTransition} is perturbation in the cost of operation with perturbation in the transition matrix and Fisher information introduced as a regularization.
	
	We consider the Total Variation distance between the rows of conditional transition matrices. Since number of states and actions are finite, the total variation distance between two discrete measures is the $l_1$-norm of the difference of the discrete distributions. So we use the $l_1$ distance between the vectorized conditional transition matrices. We employ algorithm \ref{alg:BCDTransition} which is a block co-ordinate descent algorithm to solve \eqref{eq:ECCMLogTransition}.\\
	\textbf{Remark:} We employ block co-ordinate descent to solve both \eqref{eq:ECCMLogCost} and \eqref{eq:ECCMLogTransition} because the determinant of the FIM is log-convex w.r.t $\pi$ and log-concave w.r.t $\tp(u)$ and are separable. This simplifies the optimization to apply block co-ordinate descent algorithm.
	
	
	\begin{algorithm}
		\caption{Block Co-ordinate Descent Algorithm to compute the Conditional Transition Matrix and the Optimal Policy in \eqref{eq:ECCMLogTransition}}\label{alg:BCDTransition}
		\begin{algorithmic}[1]
			\State \textbf{Input:} $\gamma_1$, $\gamma_2$, $\tp^{0}(u)$, $c_{0}(i,u)$ and $\pi_{0}(i,u)$
			\State \textbf{Initialize:} $\tp^{in}(i,u)$, $\pi_{in}(i,u)$
			\State \textbf{Define:} $\boldsymbol{\varphi}_{\pi}(\tp(u)):=\gamma_1\sum_{u}\lVert \ve(P^{T}(u))-\ve(P_{0}^{T}(u))\rVert_{1}\nonumber+\gamma_2\log\Bigg(\frac{\prod_{i\in\states}(\sum_{u}\pi(i,u))^{|\states||\actions|}}{\prod_{i,j\in\states,u\in\actions}\tp(u)}\Bigg)$
			\State \textbf{Define:} $\boldsymbol{\varphi}_{\tp(u)}(\pi):=\Bigg(\sum_{i}\sum_{u}c_0(i,u)(\pi(i,u)-\pi_{0}(i,u))\Bigg)^2+\gamma_2\log\Bigg(\frac{\prod_{i\in\states}(\sum_{u}\pi(i,u))^{|\states||\actions|}}{\prod_{i,j\in\states,u\in\actions}\tp(u)}\Bigg)$
			\State \textbf{Define:} $\mathcal{S}_{\tp(u)}=\{\pi: \quad\pi(i,u)> 0\,\text{and}\,\sum_{u}\pi(j,u)=\sum_{i}\sum_{u}\tp(u)\pi(i,u),\,j\in\states\,\text{and}\,\sum_{i}\sum_{u} \pi(i,u)=1\}$
			\State \textbf{Define:} $\mathcal{S}_{\pi}=\{\tp(u): \quad \tp(u)> 0\,\text{and}\,\sum_{u}\pi(j,u)=\sum_{i}\sum_{u}\tp(u)\pi(i,u),\,j\in\states\,\text{and}\,\sum_{j}\tp(u)=1,\,i\in\states,u\in\actions\}$
			\While{$\epsilon_{k-1}>\text{Threshold}$}
			\State $\tp^{k}(u)\gets\underset{\tp(u)\in\mathcal{S}_{\pi_{k-1}}}{\text{argmin}}\,\boldsymbol{\varphi}_{\pi_{k-1}}(\tp(u))$
			\State $\pi_k(i,u)\gets\underset{\pi\in\mathcal{S}_{\tp^{k}(u)}}{\text{argmin}}\,\boldsymbol{\varphi}_{\tp^{k}(u)}(\pi)$
			\State $\epsilon_k\gets \Vert([\ve(\tp^{k}(u));\pi_k]-[\ve(\tp^{k-1}(u));\pi_{k-1}])\Vert_{1}$
			\EndWhile\label{euclidendwhile}
			\State \textbf{return} $[\tp(u);\,\pi]$\Comment{The conditional transition matrix and the optimal policy}
		\end{algorithmic}
	\end{algorithm}
	
	
	
	\section{Example: Radar-Adversary Interaction}
	\label{Sec:RadJamInt}
	
\begin{figure}[ht!]
		\centering
		\begin{tikzpicture}
			\draw[dashed,red, thick] (-2,-1) rectangle (6,4);
			\node at (-0.5,3.75) {Slow Timescale};
			
			\draw[dashed, blue, thick] (1,-0.5) rectangle (5.5,3.5);
			\node at (3.5,-0.25) {Intermediate Timescale};
			
			\draw[dashed, green, thick] (1.5,0) rectangle (5,3);
			\node at (3.5,0.25) {Fast Timescale};
			
			\node[anchor=south west] at (-2,-1) {\includegraphics[width=2cm]{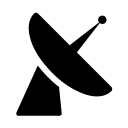}};
			\node at (-1, 1.5) {MFR};

			\node at (3.25, 2.75) {Adversaries};
			
			\node at (2.5,2) {\includegraphics[width=0.75cm]{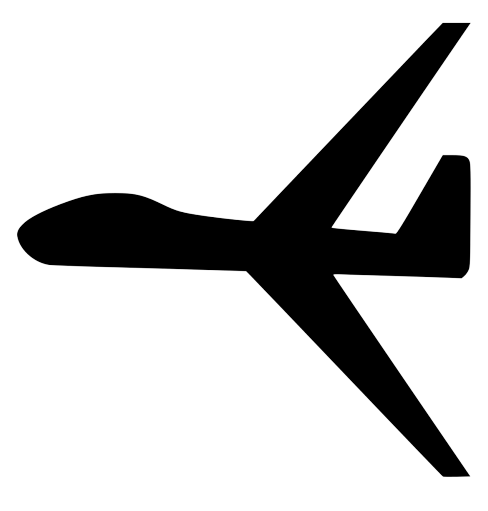}};
			
			\node at (4,2) {\includegraphics[width=1cm]{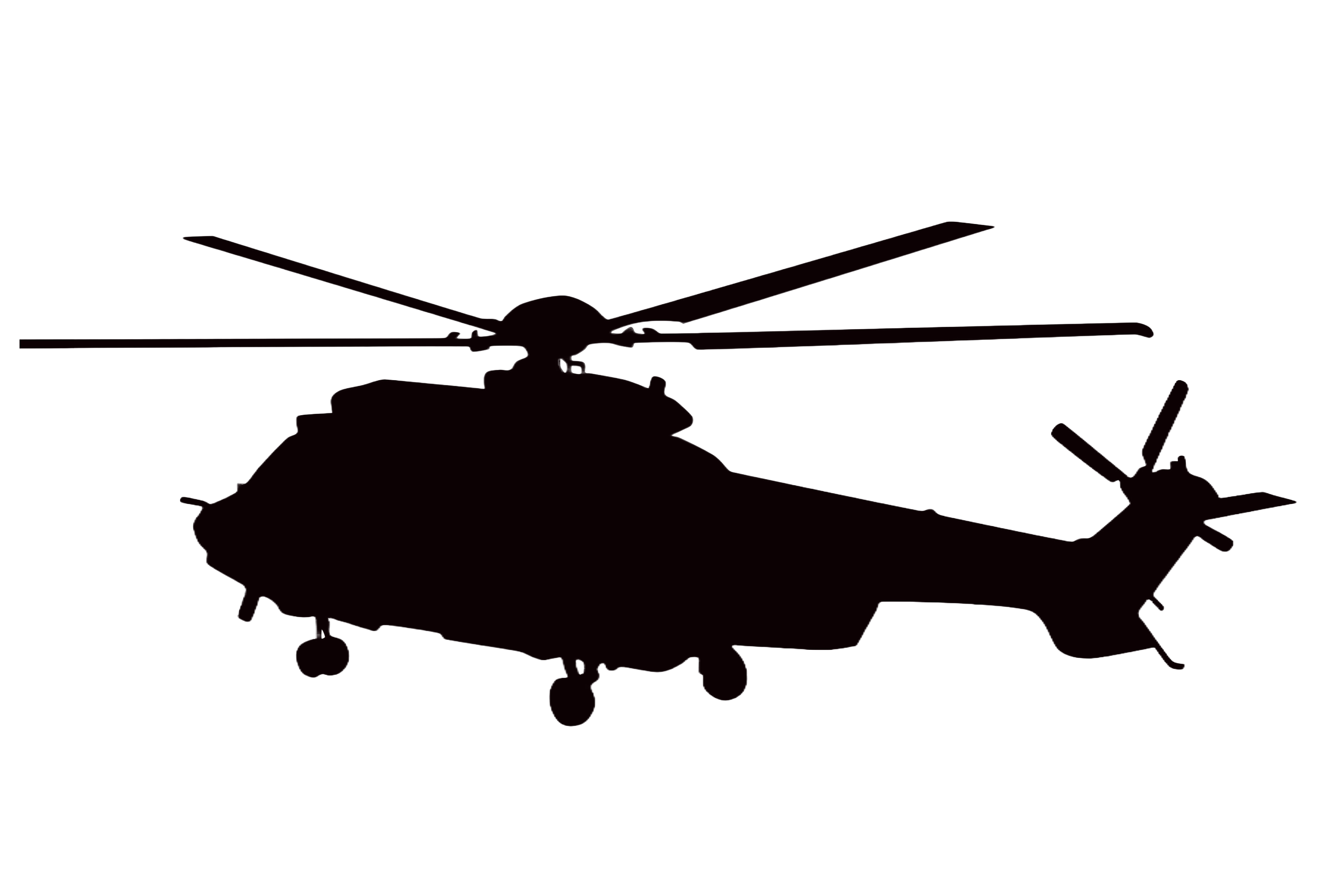}};
			
			\node at (3.5,1) {\includegraphics[width=1cm]{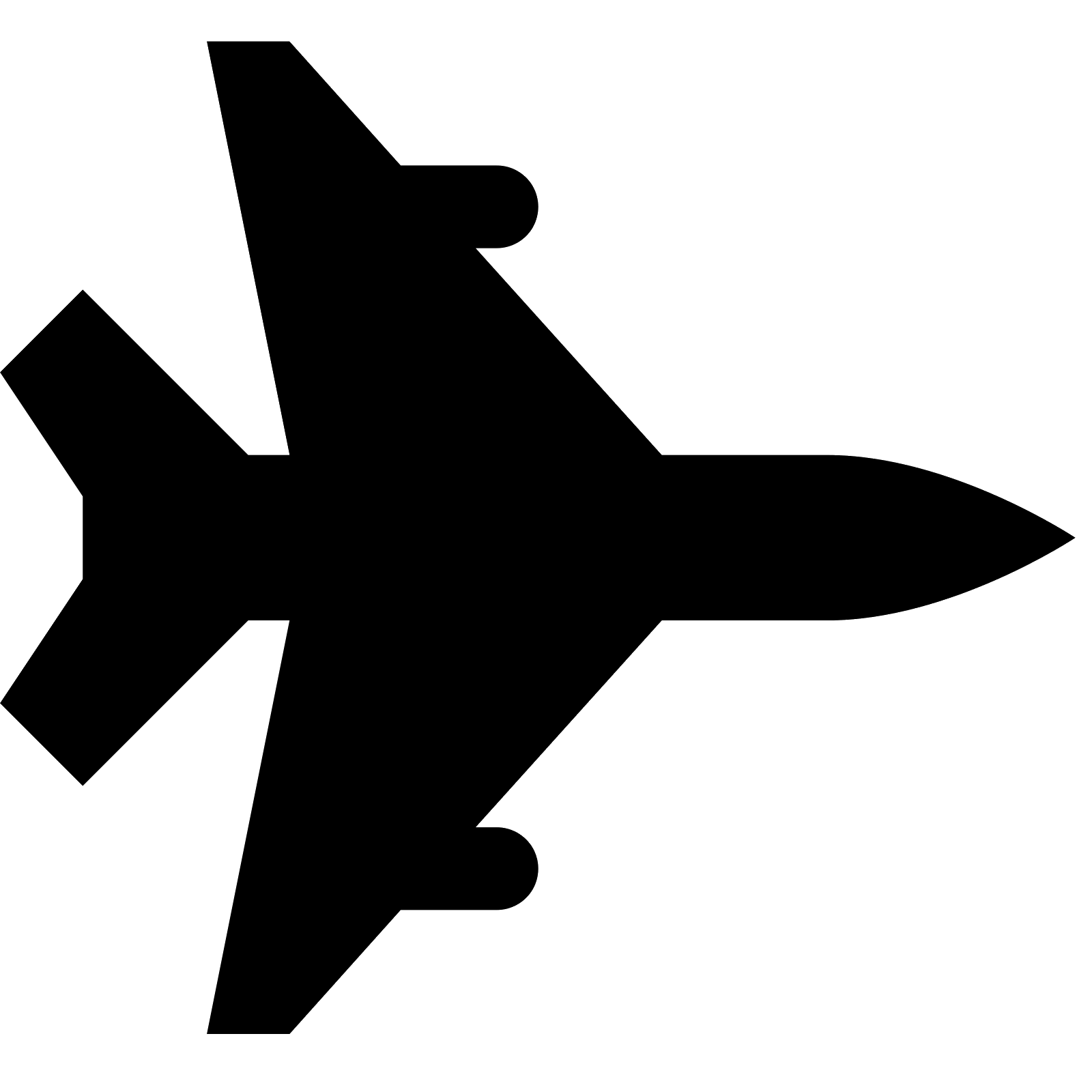}};

			\draw[->, thick, red] (0,1) -- (3,1.5);
			%
			%
		\end{tikzpicture}
		\caption{The radar (MFR) is operating in a scenario that involves multiple targets and switches through numerous modes over a slow timescale (blue arrow). The adversaries are also observing the state-action transitions and trying to infer the sensing plan. The radar's objective is to deceive the adversary by increasing the error variance in the estimate of the sensing plan (red arrow).}
		\label{Fig:ECCM}
	\end{figure}
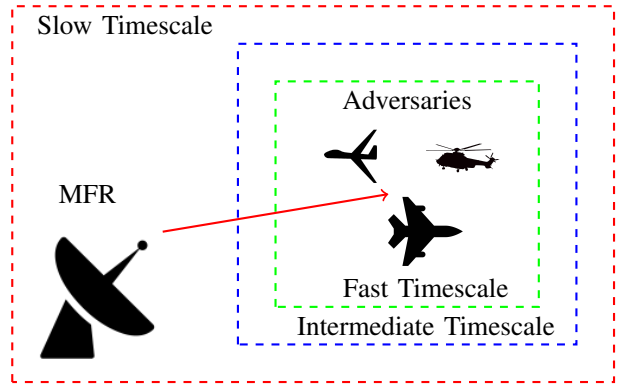

	In this section, we present the radar-adversary interaction illustrated in Fig.\,\ref{Fig:ECCM} in three time-scales. The MFR is an automated system that utilizes dynamic parameter adjustments through multi-modal switching operations to function in complex environments with multiple targets. The functions executed by the MFR are organized across hierarchical layers, spanning from pulse generation in the physical domain to meta-cognition level command and control.
	The MFR processes data over different timescales. We divide the MFR operation into three timescales: the fast timescale indexed by $l$, the intermediate timescale indexed by $q$, and the slow timescale indexed by $k$. In the fast timescale, the radar processes data at the physical layer, i.e., pulse by pulse. In the intermediate timescale, the radar processes the search and track data of the targets. In the slow timescale, the radar makes decisions for state-action transitions. The signals are interleaved between the adversary and the radar in the slow timescale and time progression is non-uniform w.r.t fast timescales. We link the parameters of the physical layer to the parameters of the control and command layer.

	\textbf{Physical Layer ($l$):} Assuming the Space Time Adaptive Processing (STAP) framework \cite{Richards2022}, the physical layer consists of data which spans across range-bins, pulses and multiple channels. We assume that the radar has a perfect estimate of the target plus clutter plus noise covariance matrix denoted as $\mathbf{R}$. $\mathbf{R}=\mathbf{R}_t+\mathbf{R}_c+\sigma^2\mathbf{I}$, where $\mathbf{R}_t$ corresponds to the target covariance matrix, $\mathbf{R}_c$ to the clutter covariance matrix and the $\sigma^2$ corresponds to the noise power. The receive SINR $\rho$ for a perfect estimate of the target plus clutter and noise covariance matrix $\mathbf{R}$ and a given scan signal $\mathbf{z}$ for a given range-bin in the STAP setup is denoted as $\rho=\mathbf{z}^{H}\mathbf{R}^{-1}\mathbf{z}$. Given that $\mathbf{R}^{-1}=\sum_{i=1}^{r_{t}}b_{i}^{t}\mathbf{v_{i}v_{i}}^{H}+\sum_{i=1}^{r_{c}}b^{c}_{i}\mathbf{v_{i}v_{i}}^{H}+\sigma^{2}\mathbf{I}$, where $r_{t}$ and $r_{c}$ correspond to the target and clutter rank for a given range bin. The terms $b_{i}^{t}$ and $b_{i}^{c}$ correspond to the target and clutter matrix singular values, with $\mathbf{v}_{i}$ denoting the eigenvectors corresponding to the index $i$. The expression for the receive SINR is $\rho=\sum_{i=1}^{r_t}b_{i}^{t}|\langle\mathbf{v}_i,\mathbf{z}\rangle|^2+\sum_{i=1}^{r_c}b_{i}^{c}|\langle\mathbf{v}_i,\mathbf{z}\rangle|^2+\sigma^2\langle\mathbf{z},\mathbf{z}\rangle$.
	The radar scans across all range bins using wideband adaptive beamforming. The scanning may be achieved using helical, spiral, raster, Palmer and Nodding patterns.  For target detection, for a fixed range-bin, the radar employs low-rank adaptive normalized matched filter  (LR-ANMF) as demonstrated in \cite{Jain2023Clutter}. In the physical layer the radar estimates the target parameters which include position and velocity.\\
	\textbf{Tracking  and Classification Layer ($q$):} In this timescale, the radar tracks a set of targets using standard filtering techniques like the Kalman filter, Extended Kalman filter and particle filter. The radar performs track maintenance for each target and classifies the targets based on their Radar Cross Section (RCS) and velocity patterns.\\
	\textbf{Command and Control Layer ($k$):} In this layer, the radar switches between various modes of operations depending upon the costs. The switching is modelled using an MDP, which we defined in Sec.\,\ref{Sec:GO}-\ref{Sec:MDP}.
	The states $x_k$ are characterized by the parameters like transmit power, pulse length, PRF, bandwidth, modulation, polarization, and dwell time. The radar's actions $u_k$ encompass tasks such as target search and tracking. Since the radar has a fixed receive SINR range and is using finite state Markov chain, $\rho$ is discretized uniformly into $|\states|$ bins. Within each $i^{\text{th}}$ bin, each state represents the SINR range which the radar is receiving. The transition probability $\tp(u)$ is a function of the SINR and the action. For the same receive SINR, the radar may switch from target tracking to scanning mode in order to acquire information on additional targets in the scenario depending on the sensing plan. We present a model for the state-action cost and the conditional transition matrix based on \cite{Selvi2018}.\\
	\textbf{Cost Function}:
	The cost of action $u_k$ given a state $x_{k}$ depends on the current SINR $\rho$ pertaining to the state, and the cost of the action. We model the state-action cost incurred by the radar in the absence of the adversary by the following expression:
	\begin{equation}
		\label{eq:Cost}
		c_0(i,u_k)=C_{\rho_i}C_{u_k}
	\end{equation}
	where $C_\rho$ is a monotonically decreasing function w.r.t $\rho$ and $C_{u}$ is an action dependent operation cost which the radar incurs for taking a particular action. $C_{\rho_k}$ should be monotonic as a higher SINR guarantees a lower cost of operation for the radar, we model $C_\rho=1-\tanh(\rho_{\text{dB}}/\chi)$, where $\rho_{\text{dB}}$ is SINR in dB and $\chi$ is a system dependent constant which represents the rate at which the costs vary with the states. The term $C_{u_k}$ is a system dependent term, and depends on the processing power and processing time for an action.\\
	\textbf{ Conditional Transition Matrix}: For scanning based actions, the radar is searching for potential targets, the conditional transition matrix depends upon the SINR of the current state and the subsequent state. Note that since targets in lower SINR regions are harder to detect, the radar would spend more time in lower SINR states than the higher SINR states. For tracking based actions, the radar tracks different sets of targets spread across multiple SINR states. The transition probabilities are dependent upon the SINR states corresponding to the targets and like searching, the radar would spend more time in low SINR states. We model the conditional transition matrix in the absence of adversary using the following expression:
	\begin{equation}
		\label{eq:TransProb}
		\tp^{0}(u)=p(\rho_i-\rho_j,u)=\frac{\exp(K_{i}t_u(\rho_i-\rho_j)_{\text{dB}})}{\sum_{j\in\states}{\exp(K_{i}t_u(\rho_i-\rho_j)_{\text{dB}})}}
	\end{equation}
	such that for a given action $u$, $p(\rho_i-\rho_s,u)\geq p(\rho_{i}-\rho_{j},u),\,j\geq k, i,\,j,\,k\in\states$ and $\sum_{j\in\states}\tp(u)=1$. The function $t_u$ is action dependent like $C_{u}$, and depends upon the processing time and power. $K_i$ an SINR dependent term that influences the transition and accommodates imperfect SINR measurement and increases with SINR. Observe that the radar has higher probability to stay in the lower SINR states, the cost of operating in low SINR states is high, so the radar needs to optimally design its policy so that cost of operation minimized. The state-action costs and conditional transition matrix abstract the parameters from the physical layer to the command and control layer.
	
	An adversary observes radar and infers actions and the radar's states by interleaving the signals in the slow timescale to obtain $\history_k$, recall that $\history_k$ is the sequence of states and actions $\{x_0u_0\dots u_{k-1}x_k\}$. The data $\history_k$ is sampled from $\mu$, where $\mu$ represents the sensing plan. The adversary uses an unbiased MLE, \eqref{eq:MLE}, to estimate the sensing plan by observing the state-action trajectory.
	
	In the absence of  the adversary the radar devises the sensing plan by solving \eqref{eq:LPPrim} with costs defined in \eqref{eq:Cost} and conditional transition matrix defined in \eqref{eq:TransProb}. When the adversary is present the radar aims to devise a masking strategy such that the variance in the error of the estimate is large. The radar needs to increase the determinant of the covariance matrix, thereby decreasing the determinant of the FIM of the adversary which can be achieved by radar devising a sensing plan by either solving \eqref{eq:ECCMLog1}, or \eqref{eq:ECCMLogCost}, or \eqref{eq:ECCMLogTransition}, depending upon the operational requirement.

	\section{Numerical Results}
	\label{Sec:Num}
	We show numerically that determinant of the Fisher information matrix is reduced in cases when we perturb: a) the total cost of operation (Sec.\,\ref{Sec:Num}-\ref{Sec:RandDet}), b) the state-action costs (Sec.\,\ref{Sec:Num}-\ref{Sec:StateActionNum}) and c) the conditional transition matrix (Sec.\,\ref{Sec:Num}-\ref{Sec:TransMatNum}). In Sec.\,\ref{Sec:Num}-\ref{Sec:EmpError}, we empirically demonstrate that the Fisher Information yields a higher error in the estimate of the conditional transition matrix  compared to the Maximum Entropy criteria\,\cite{Savas2020}.
	\subsection{Masking Plan by Perturbing the Total Operation Cost}
	\label{Sec:RandDet}
\begin{figure}
		\centering
		\includegraphics[width=\linewidth]{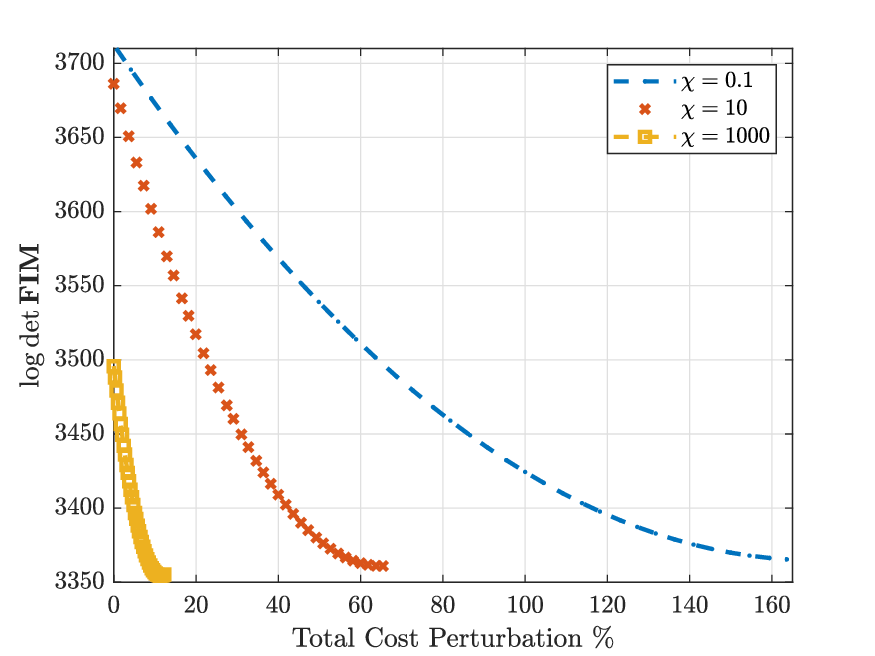}
		\caption{The perturbation in the total operation cost and Fisher information are inversely related. The parameter $\chi$ quantifies the rate of variation of instantaneous costs across states for a given action. Lower the $\chi$, higher the variation of the instantaneous costs across the states. For a given decrease in the adversary's Fisher information for the estimate of the sensing plan, the total operation cost increases as the rate of variation of the state-action cost across states decreases.}
		\label{fig:FIMvPert}
	\end{figure}

	We solved the optimization problem \eqref{eq:ECCMLog1} using the toolbox \texttt{fmincon} in Matlab for 10 states, 4 actions. The SINR range is from 0dB to 35dB and the actions denote $\{$Fine Scanning, Coarse Scanning, Fine Tracking, Coarse Tracking$\}$, for the state-action costs $c_{0}(i,u)$ as modelled in \eqref{eq:Cost}, the parameter $C_u=\{0.606,\,0.407,\, 0.977,\, 0.465\}$. For the conditional transition matrix $\tp(u)$ as modelled in \eqref{eq:TransProb}, the parameter $K_{i}=\{$0.0040, 0.0210, 0.0960, 0.1310, 0.2130, 0.5020, 0.5280, 0.7910, 0.8450, 0.8500$\}$ and $t_{u}=\{0.083,\,0.413,\, 0.590,\, 0.928\}$\footnote{The values of $K_i$, $C_{u}$ and $t_u$ have been randomly defined, however the values satisfy the criteria defined in Sec.\,\ref{Sec:RadJamInt}. For a given MFR, $C_{u}$ is a processing power dependant term. $K_i$ and $t_u$, depend on a the stationary clutter related parameters of a scenario which can be estimated for each SINR range by operating the radar when no targets are present.}.
	Since \eqref{eq:ECCMLog1} is a nonlinear optimization problem, we set a random initial point and averaged the policy over 200 Monte Carlo simulations for each Lagrange multiplier value $\gamma$, swept from $10^{-3}$ to $10^{-1}$ and evaluated the $\log\det\FIM$ and total cost perturbation at the optimal value $\mu_{*}$ in \eqref{eq:ECCMLog1}. In Fig.\,\ref{fig:FIMvPert} we demonstrate that the determinant of the FIM and the perturbation of the operation cost for the radar are inversely related. We plot the inverse relation between FIM and the total operation cost for different rates of instantaneous cost variation across the states for a given action, quantified by $\chi$ in \eqref{eq:Cost}. In Table \ref{Table:RandomPolicy}, we illustrate the masked sensing plan, with $\chi=10$, at around 20$\%$ perturbation in the total operation where rows in the action columns represent the probability $u_k\sim\mu(\rho_i)$. Observe that at such high perturbation in the operation cost, the radar has more probability for coarse tracking at all states. 
	
	\begin{table}
		\caption{Masked Sensing Plan}
		\begin{tabular}{|p{15mm}|p{10mm}|p{10mm}|p{10mm}|p{10mm}|}
			\hline
			{SINR (dB)}&     Fine Scanning   &    Coarse Scanning    &   Fine Tracking     &   Coarse Tracking     \\\hline
			0-3.5 & 0.0090 & 0.0309 & 0.0061 & 0.9539 \\\hline
			3.5-7 & 0.0214 & 0.0330 & 0.0212 & 0.9245 \\\hline
			7-10.5 & 0.0384 & 0.0492 & 0.0429 & 0.8695 \\\hline
			10.5-14 & 0.0586 & 0.0602 & 0.0568 & 0.8244 \\\hline
			14-17.5 & 0.0657 & 0.0701 & 0.0684 & 0.7958 \\\hline
			17.5-21 & 0.0745 & 0.0778 & 0.0770 & 0.7707 \\\hline
			21-24.5 & 0.0833 & 0.0842 & 0.0843 & 0.7481 \\\hline
			24.5-28 & 0.0875 & 0.0919 & 0.0919 & 0.7287 \\\hline
			28.31.5 & 0.0932 & 0.0956 & 0.0966 & 0.7146 \\\hline
			31.5-35 & 0.0975 & 0.0998 & 0.0997 & 0.7031\\\hline
		\end{tabular}
		\label{Table:RandomPolicy}
	\end{table}
	\subsection{Masking Plan by Perturbing the State-Action Cost}
	\label{Sec:StateActionNum}
	
\begin{figure}
		\centering
		\begin{subfigure}[b]{0.85\linewidth}
			\centering
			\includegraphics[width=\textwidth]{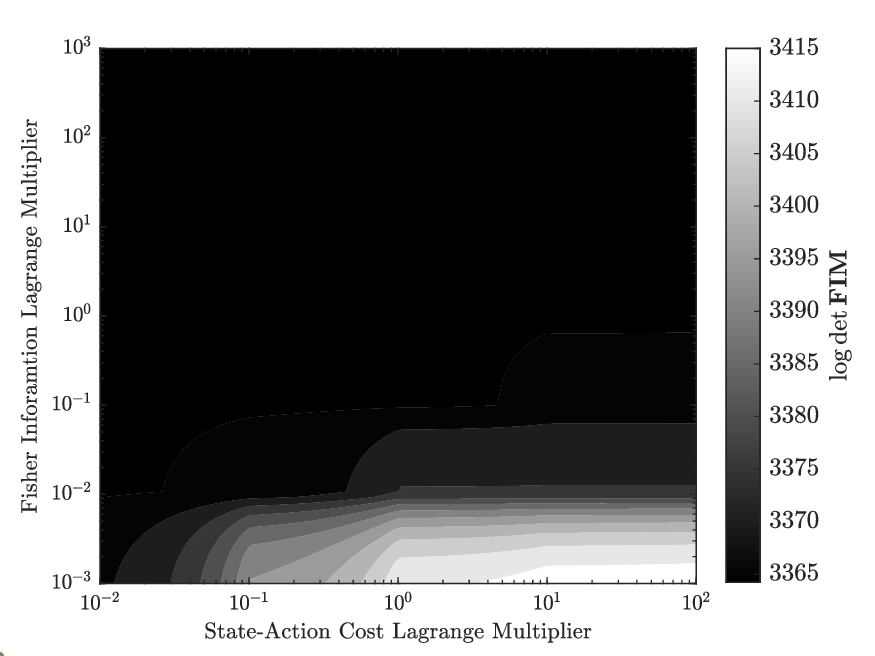}\subcaption{}
			\label{fig:FIMvsGamma2}
		\end{subfigure}\hfill
		\begin{subfigure}[b]{0.85\linewidth}
			\centering
			\includegraphics[width=\textwidth]{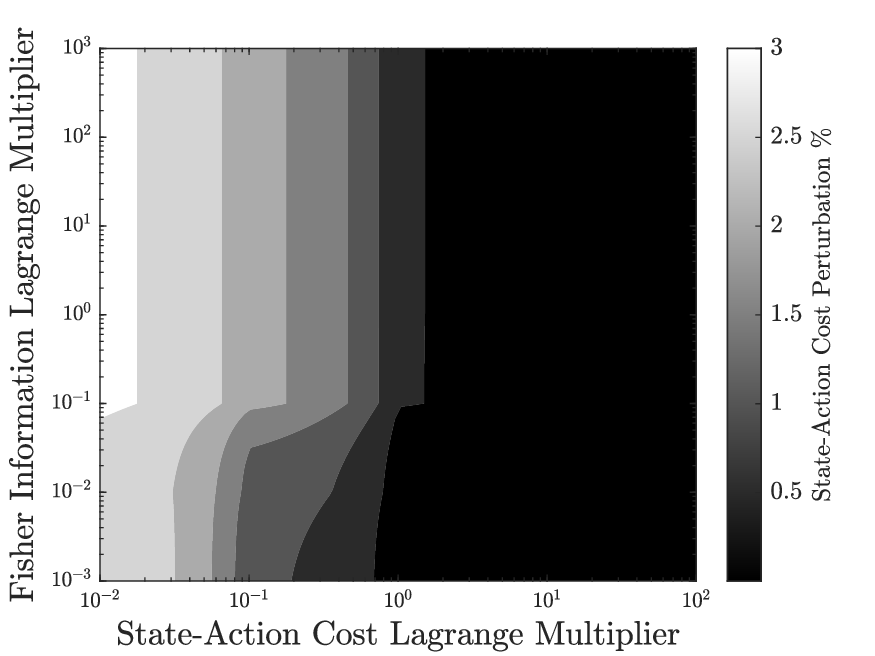}\subcaption{}
			\label{fig:StateActionCostvsGamma1}
		\end{subfigure}\hfill
		\begin{subfigure}[b]{0.85\linewidth}
			\centering
			\includegraphics[width=\textwidth]{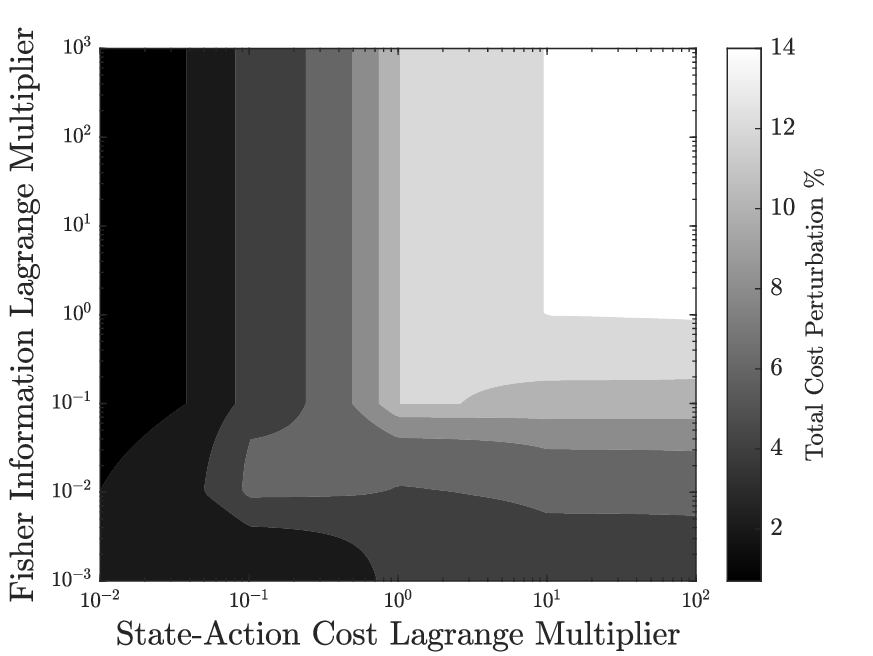}\subcaption{}
			\label{fig:TcostvsRegu}
		\end{subfigure}
		\caption{If the radar controller emphasizes on the Fisher information constraint by increasing the Fisher information Lagrange multiplier; the determinant of the FIM decreases, Fig.\,\ref{fig:FIMvsGamma2}, the perturbation in the state-action cost increases, Fig.\,\ref{fig:StateActionCostvsGamma1}. By increasing the Lagrange multiplier on the state-action costs; the determinant of the FIM increases, Fig.\,\ref{fig:FIMvsGamma2}, the perturbation in the state-action costs decreases, Fig.\,\ref{fig:StateActionCostvsGamma1}.  If the radar  emphasises more on minimizing the state-action cost perturbation and reducing the adversary's FIM then it will have an increased total cost of operation, Fig.\,\ref{fig:TcostvsRegu}.}
		\label{Fig:StateAction}
	\end{figure}

	For a 10-state-4-action example we assigned $\tp^{0}(u)$ and $c_0(i,u)$ for the given MDP with values used in the previous subsection, and computed $\pi_0$ by solving \eqref{eq:LPPrim}. We randomly initialized $c_{in}$ and $\pi_{in}$, 
	line 7 and line 8 of algorithm \ref{alg:BCD} were evaluated using \texttt{cvx} and \texttt{fmincon} from Matlab, respectively. We simulated 200 Monte Carlo simulations with random initialization. The state-action Lagrange multiplier, $\gamma_1$, and the Fisher information Lagrange multiplier, $\gamma_2$, were swept from $10^{-2}$ to $10^{2}$ and $10^{-3}$ to $10^{3}$, respectively. Fig.\,\ref{fig:FIMvsGamma2}, Fig.\,\ref{fig:StateActionCostvsGamma1} and Fig.\,\ref{fig:TcostvsRegu} demonstrate that the adversary's Fisher Information, radar's state-action cost perturbation and total perturbation in the cost of operation are inversely related. The figures serve as lookup tables for the controller tuning radar parameters. For instance, to minimize perturbations in total operational cost while simultaneously reducing the adversary's Fisher Information, an increase in state-action cost perturbations is observed. 
	
	\subsection{Masking Plan by Perturbing the Conditional Transition Matrix}
	\label{Sec:TransMatNum}
\begin{figure}
		\centering
		\begin{subfigure}[b]{0.85\linewidth}
			\centering
			\includegraphics[width=\linewidth]{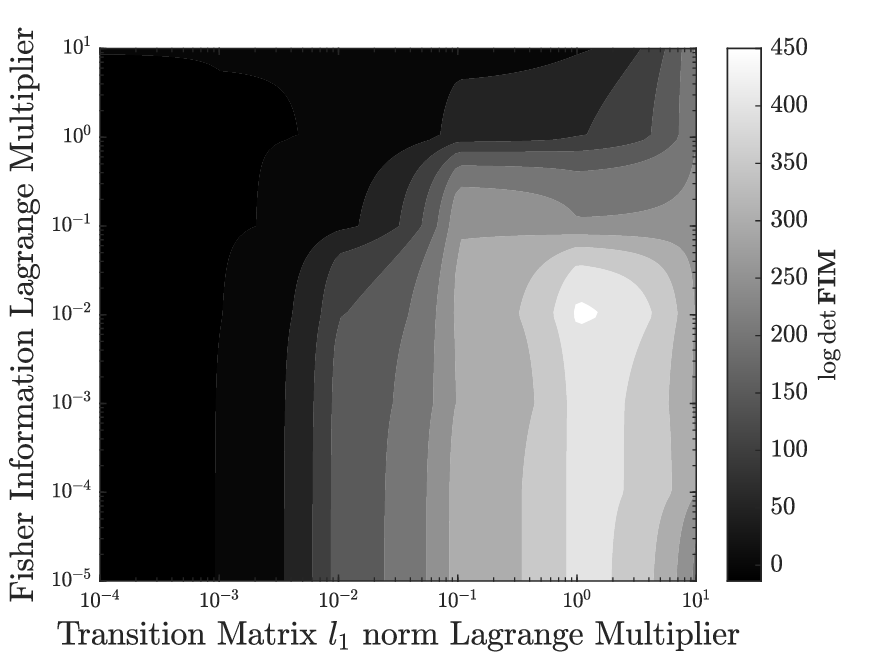}
			\subcaption{}
			\label{fig:detFIMTransProb}
		\end{subfigure}
		\hfill
		\begin{subfigure}[b]{0.85\linewidth}
			\centering
			\includegraphics[width=\linewidth]{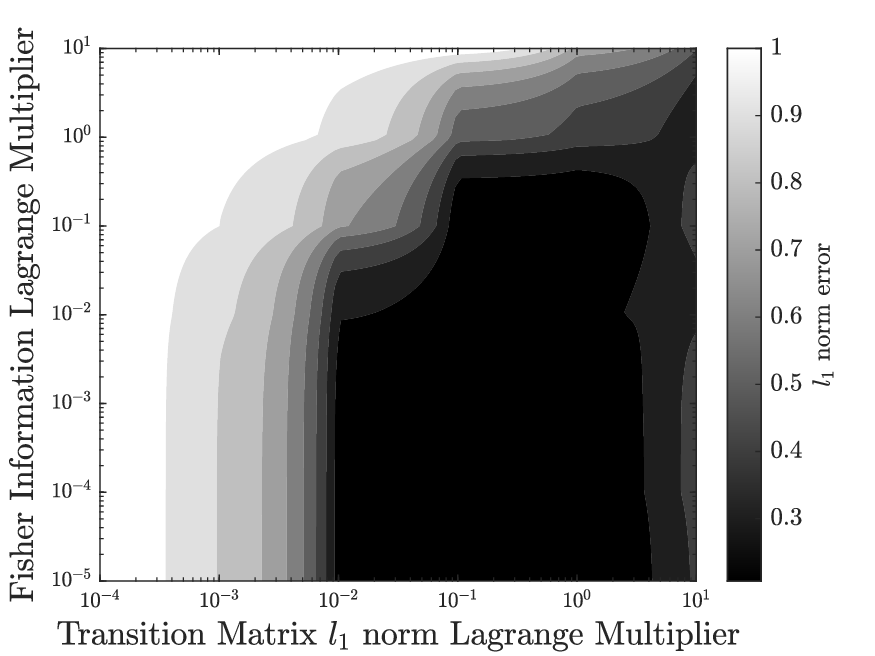}
			\subcaption{}
			\label{fig:TransProbPert}
		\end{subfigure}
		\hfill
		\begin{subfigure}[b]{0.85\linewidth}
			\centering
			\includegraphics[width=\linewidth]{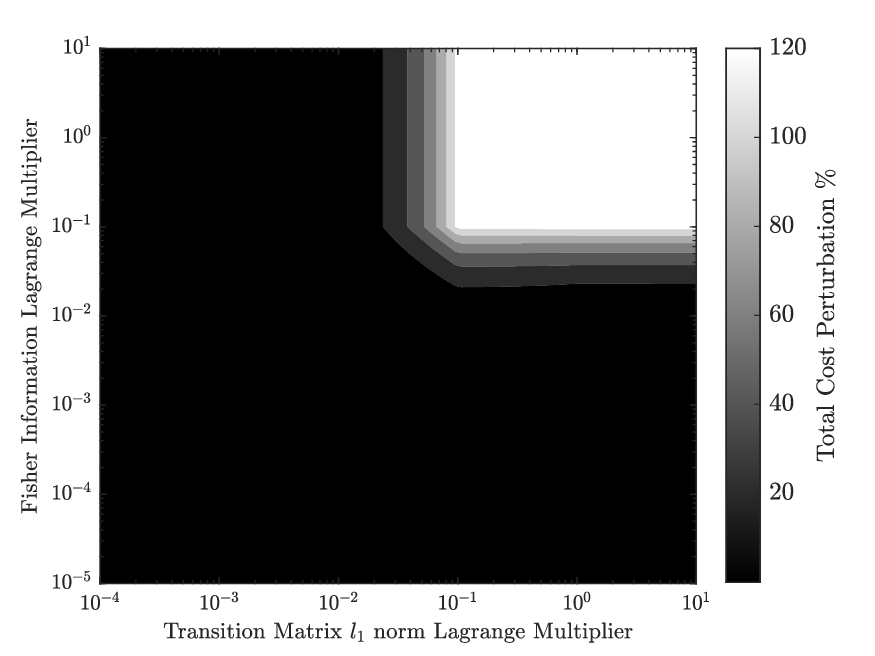}
			\subcaption{}
			\label{fig:TotalCostTransProb}
		\end{subfigure}
		\label{Fig:TransMat}
		\caption{If the radar controller emphasizes on the Fisher information constraint by increasing the Fisher information Lagrange multiplier; the determinant of the FIM decreases, Fig.\,\ref{fig:detFIMTransProb}, the perturbation in the conditional transition matrix increases, Fig.\,\ref{fig:TransProbPert}. By increasing the Lagrange multiplier on the conditional transition matrix perturbation; the determinant of the FIM increases, Fig.\,\ref{fig:detFIMTransProb}, the perturbation in the conditional transition matrix decreases, Fig.\,\ref{fig:TransProbPert}.  If the radar  emphasises more on minimizing the transition matrix perturbation and reducing the adversary's FIM then it will have an increased total cost of operation, Fig.\,\ref{fig:TotalCostTransProb}.}
		
	\end{figure}

	For a 10-states, 4-actions example we assigned $c_0(i,u)$ and $\tp^{0}(u)$ same values as used in the previous subsection. We randomly initialized $\pi_{in}(i,u)$ and $\tp^{in}(u)$ in line 2 of algorithm \ref{alg:BCDTransition}. We swept The state-action Lagrange multiplier, $\gamma_1$, and the Fisher information Lagrange multiplier, $\gamma_2$, from $10^{-4}$ to $10^{1}$ and $10^{-5}$ to $10^{1}$, respectively. Line 8 and line 9 were evaluated using the function \texttt{fmincon} using the built in interior point method. We were able to show local convergence in Fig.\,\ref{fig:detFIMTransProb}, \ref{fig:TransProbPert} and \ref{fig:TotalCostTransProb}. We demonstrated that reduction in the determinant of the Fisher information matrix and the transition matrix perturbation were inversely related for a fixed perturbation in the total cost of operation. As in the previous section, the figures represent a look-up table which the operator can use to tune the parameters to achieve the desired Fisher information.
	\subsection{Masking of Transition Logic for the adversary}
	\label{Sec:EmpError}
\begin{figure}
		\centering
		\includegraphics[width=\linewidth]{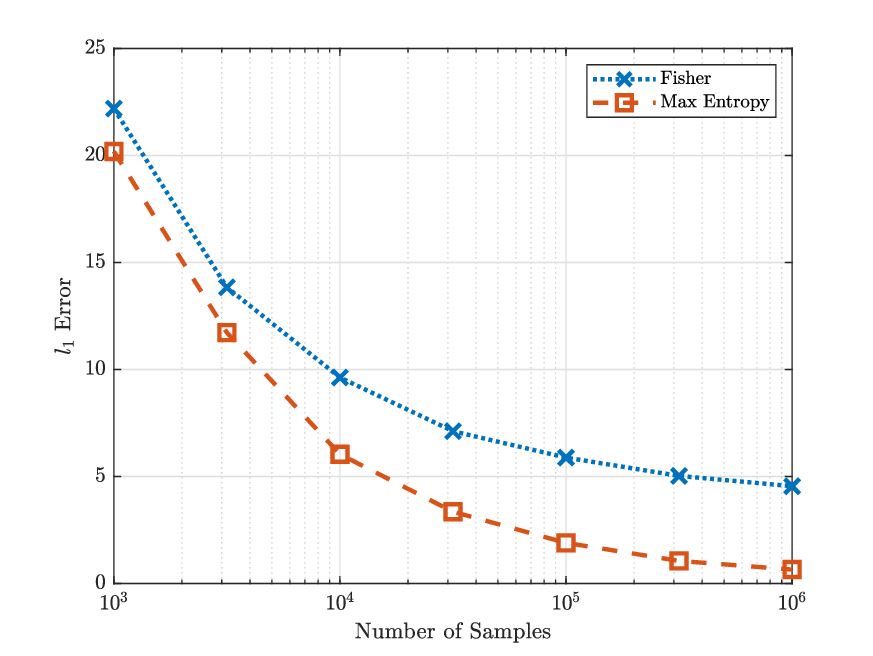}
		\caption{The Fisher information criteria outperforms the Maximum Entropy method in concealing the conditional transition matrix, as demonstrated by the sum of $l_1$ distances between the adversary's estimates (computed using Fisher information and Maximum Entropy criteria) and the true conditional transition matrix. Higher efficacy is indicated by the larger error in the estimate of the conditional transition matrix using the Fisher information criteria compared to the Maximum Entropy criteria.}
		\label{fig:Comparison}
	\end{figure}

	In this subsection, we empirically compare the efficacy of the Fisher information criteria and the Maximum Entropy criteria in concealing the conditional transition matrix of the radar controller. The Maximum Entropy criteria, as mentioned in \cite{Savas2020}, incorporates the constraint $-\sum_{i\in\states,u\in\actions}\pi(i,u)\log\,\pi(i,u)$ in \eqref{eq:ECCMLog1}, replacing the Fisher information constraint. Our scenario involves an adversary observing the radar controller's noisy transitions, where 'noise' refers to the adversary observing finite samples. We evaluate the adversary's estimate of the conditional transition matrix concerning the original matrix used by the radar controller (defined in Sec.\,\ref{Sec:Num}-\ref{Sec:RandDet}) using both Fisher information and Maximum Entropy criteria.  To quantify the performance, we compute the total variation distance between the rows of the estimated conditional transition matrix and the true matrix for all actions. The results, illustrated in Fig.\,\ref{fig:Comparison}, demonstrate that the sum of the total variation distance for the Fisher information criteria is higher than that for the Maximum Entropy criteria from the true conditional transition matrix.
	\section{Conclusion}
	\label{Sec:Conclusion}
	We have demonstrated that an adversary's accuracy in estimating the sensing plan of an MDP-based radar controller can be degraded using the Fisher information criteria. Furthermore, our demonstration revealed that perturbations in the total cost of operation, state-action costs, and conditional transition matrix led to a reduction in the determinant of the Fisher information available to the adversary. The proposed method yields a higher error in the estimate of the sensing plan compared to the sensing plan devised using maximum entropy criteria.
	
	As a natural extension of our current work, we aim to extend the application of Fisher information criteria to an SCFG-based MFR controller. Acknowledging the asymptotic nature of the Fisher information criteria, we plan to propose a finite sample criteria by incorporating concepts from differential privacy, a methodology applicable to both MDP and SCFG-based radar controllers. 
	
	%
	
	\bibliography{Bibliography.bib}
	\bibliographystyle{IEEEtran}
	
\end{document}